\documentclass[mathleft]{an_mod2}
\usepackage{graphicx}
\usepackage{times}
\usepackage{amssymb}
\usepackage{natbib}
\bibpunct{(}{)}{;}{a}{}{,} 

\def\chandra{{\it Chandra~}}
\def\chandrak{{\it Chandra}}
\def\swift{{\it Swift~}}
\def\swiftk{{\it Swift}}
\def\xmm{{XMM-Newton~}}
\def\xmmk{{XMM-Newton}}

\def\m31{{M~31}}
\def\msun{{$M_{\sun}$}}

\newcommand{\nh}{\hbox{$N_{\rm H}$}~}
\newcommand{\hcm}[1]{$\times 10^{#1}$ cm$^{-2}$}
\newcommand{\ohcm}[1]{$10^{#1}$ cm$^{-2}$}

\newcommand{\oergs}[1]{$10^{#1}$ erg s$^{-1}$}

\newcommand{\power}[1]{$10^{#1}$}

\def\novag{{M31N~2007-06b~}}
\def\novagk{{M31N~2007-06b}}
\def\gsss{{1E~1339.8+2837~}}

\def\novaf{{M31N~2007-11a~}}
\def\novafk{{M31N~2007-11a}}

\def\novap{{M31N~2006-04a~}}
\def\novapk{{M31N~2006-04a}}

\begin{document}

\Pagespan{789}{}
\Yearpublication{2010}%
\Yearsubmission{2009}%
\Month{07}%
\Volume{}%
\Issue{}%

\title{Recent discoveries of supersoft X-ray sources in \m31\thanks{Partly 
   based on observations with \xmmk, an ESA Science Mission with instruments and contributions directly funded by ESA Member States and NASA}}

\author{M. Henze\inst{1}\fnmsep\thanks{Corresponding author: \email{mhenze@mpe.mpg.de}\newline}
	\and  W.~Pietsch\inst{1}
	\and F.~Haberl\inst{1}
	\and G.~Sala\inst{1,2}
	\and M.~Hernanz\inst{3}
	\and D.~Hatzidimitriou\inst{4,5}
	\and A.~Rau\inst{1,6}
	\and D.H.~Hartmann\inst{7}
	\and J.~Greiner\inst{1}
	\and M.~Orio\inst{8}	
	\and H.~Stiele\inst{1}
	\and M.J.~Freyberg\inst{1}
}

\titlerunning{Recent discoveries of SSSs in \m31}

\authorrunning{M. Henze et al.}
\institute{Max-Planck-Institut f\"ur extraterrestrische Physik, D-85748 Garching, Germany
	\and Departament de F\'isica i Enginyeria Nuclear, EUETIB (UPC/IEEC), Comte d'Urgell 187, 08036 Barcelona, Spain
	\and Institut de Ci\`encies de l'Espai (CSIC-IEEC), Campus UAB, Fac. Ci\`encies, E-08193 Bellaterra, Spain
	\and Department of Astrophysics, Astronomy and Mechanics, Faculty of Physics, University of Athens, Panepistimiopolis, GR15784 Zografos, Athens, Greece
	\and Foundation for Research and Technology Hellas, IESL, Greece
	\and California Institute of Technology, Pasadena, CA 91125, USA
	\and Department of Physics and Astronomy, Clemson University, Clemson, SC 29634-0978, USA
	\and INAF, Osservatorio Astronomico di Padova, Vicolo dell' Osservatorio, I-35122 Padova, Italy 
}

\received{31 July 2009}
\accepted{02 December 2009}
\publonline{later}

\keywords{galaxies: individual: (\m31) -- novae, cataclysmic variables -- stars: individual: (\novagk, \novafk, \novapk) -- X-rays: binaries}

\abstract{
Classical novae (CNe) have recently been reported to represent the major class of supersoft X-ray sources (SSSs) in the central area of our neighbouring galaxy \m31. This paper presents a review of results from recent X-ray observations of \m31 with \xmm and \chandrak. We carried out a dedicated optical and X-ray monitoring program of CNe and SSSs in the central area of \m31. We discovered the first SSSs in \m31 globular clusters (GCs) and their connection to the very first discovered CN in a \m31 GC. This result may have an impact on the CN rate in GCs. Furthermore, in our optical and X-ray monitoring data we discovered the CN \novafk, which shows a very short SSS phase of 29 -- 52 days. Short SSS states (durations $\le $ 100 days) of CNe indicate massive white dwarfs (WDs) that are candidate progenitors of supernovae type Ia. In the case of \novafk, the optical and X-ray light curves suggest a binary containing a WD with $M_{WD} >$ 1.0 \msun. Finally, we present the discovery of the SSS counterpart of the CN \novapk. The X-ray light curve of \novap shows short-time variability, which might indicate an orbital period of about 2 hours.
}

\maketitle

%
%
\section{Introduction}
%
The class of supersoft X-ray sources (SSSs) was first characterised on the basis of ROSAT observations \citep[e.g.][]{1991A&A...246L..17G}. These sources show extremely soft X-ray spectra, with little or no emission at energies above 1 keV \citep[see e.g.][]{1998A&A...332..199P}, that can be described by equivalent black body temperatures of $\sim$15-80 eV \citep[see][and references therein]{1997ARA&A..35...69K}. \citet{2005A&A...442..879P} found that classical novae (CNe) represent the major class of SSSs in the central area of \m31. They showed that more than 60\% of the SSSs from the \xmm survey of \m31 by \citet{2005A&A...434..483P} can be identified with novae.

CNe originate in thermonuclear explosions on the surface of white dwarfs (WDs) in cataclysmic binaries. These explosions are resulting from the transfer of matter from the companion star to the WD. The transferred hydrogen-rich matter accumulates on the surface of the WD until hydrogen ignition starts a thermonuclear runaway in the degenerate matter of the WD envelope. The resulting expansion of the hot envelope causes the brightness of the WD to rise by more than nine magnitudes within a few days, and leads to ejection of mass at high velocities \citep[see][and references therein]{2005ASPC..330..265H,1995cvs..book.....W}. However, a fraction of the hot envelope can continue burning hydrogen steadily on the surface of the WD \citep{1974ApJS...28..247S,2005A&A...439.1061S}, powering a luminous ($L_x\sim$\oergs{37..38}), transient SSS that can be observed directly once the ejected envelope becomes sufficiently transparent for X-rays \citep{1989clno.conf...39S,2002AIPC..637..345K}.

The duration of the SSS phase of CNe is related to the amount of H-rich matter that is {\it not} ejected during the nova outburst and also depends on the mass of the WD. More massive WDs need to accrete less matter to initiate the thermonuclear runaway, because of their higher surface gravity \citep{1998ApJ...494..680J}. As a consequence, post-nova WD envelopes are smaller for more massive WDs, although this also depends on the accretion rate. Thus, the duration of the SSS state is inversely related to the mass of the WD \citep{2005A&A...439.1061S,1998ApJ...503..381T}. In turn, the time of appearance of the SSS is determined by the fraction of mass ejected in the outburst \citep{2006ApJS..167...59H}. Typically, SSS states of CNe last from months to several years \citep{2007A&A...465..375P}.

We carried out a dedicated optical and X-ray monitoring program of CNe and SSSs in the central area of our neighbour galaxy \m31 \citep[distance 780 kpc,][used throughout the paper]{1998AJ....115.1916H,1998ApJ...503L.131S}. The X-ray observations were obtained with \xmm and \chandra in three monitoring campaigns from June 2006 till March 2007 (AO5), November 2007 till February 2008 (AO6), and November 2008 till February 2009 (AO7). The AO6 and AO7 monitoring consisted of single observations separated by 10 days, in contrast to AO5 where the separation between the observations was about 50 days. We revised the monitoring strategy to be able to detect the CNe with short SSS phases as found by \citet{2007A&A...465..375P}.

In this work we discuss newly discovered objects from three different, peculiar classes of SSSs: (a) SSSs in globular clusters (GCs) in Sec.\ref{sec:gc_sss}, (b) CNe with very short SSSs phases in Sec.\,\ref{sec:novaf}, and (c) SSSs with indications for light curve periodicity in Sec.\,\ref{sec:novap}. In Sec.\,\ref{sec:discuss} we conclude with a brief summary.

%
%
\section{First supersoft X-ray sources in \m31 globular clusters}
\label{sec:gc_sss}
%
In our AO6 monitoring data we discovered the very first SSSs in \m31 globular clusters (GCs) \citep{2009A&A...500..769H}. The sources were found in the GCs Bol 111 (source SS1) and Bol 194 (source SS2) in a \chandra observation starting at 2007-11-07.64 UT (ObsID 8526). Table\,\ref{table:gc_sss} summarises the source properties. The X-ray positions of both SSSs are in good agreement with the optical positions of the two GCs. Since both sources were at large off-axis angles in the HRC-I field of view, we computed their positions using \xmm observations for SS1 and \swift observations for SS2.

To perform spectral analysis of SS1 we used \xmm observations obtained at 2008-01-05.99 UT and at 2008-02-09.31 UT (ObsIDs 0511380201 and 0511380601). Since SS2 faded before the start of the \xmm observations, we used a \swift follow-up observation (ObsID 00031027001) to constrain the spectrum. Source parameters derived from XSPEC black body fits to the spectra are shown in Table\,\ref{table:gc_sss}. The table also shows SS1 parameters derived from \swift observation 00031071002 for comparison. Note, that this observation was performed on 2007-11-19.27 UT, about 50 days before the \xmm observations, and therefore indicates a trend for some parameters. Despite the errors for the black body temperature of SS2 are large, due to the few counts in the \swift XRT spectrum, this source only emits photons with energies below 750 eV. Therefore, we classify both sources as SSSs.

SSSs in GCs are extremely rare. There was just one previously known object: the transient \gsss in the Galactic GC M\,3 \citep[NGC 5272;][]{1999PASJ...51..519D,1995A&A...300..732V}. \citet{1999PASJ...51..519D} measured for this source a black body temperature of $kT \sim$ 36 eV and a luminosity of $\sim$ \oergs{35}, which is significantly lower than the observed peak luminosities of other SSSs, including the two sources discussed here. \citet{1999PASJ...51..519D} discuss \gsss as a cataclysmic variable (CV) system that may be a dwarf nova involving a massive WD.

%
\begin{table}[ht]
\caption{Features of the SSSs in GCs. X-ray parameters of SS1 are based on \xmm (\swiftk) observations, X-ray parameters of SS2 are based on \swiftk.}
\label{table:gc_sss}
\begin{center}
\begin{tabular}{lrr}\hline\hline \noalign{\smallskip}
	  & SS1 (Bol 111) & SS2 (Bol 194)\\ \hline \noalign{\smallskip}
	RA (J2000) & 00:42:33.21 & 00:43:45.30\\
	Dec (J2000) & +41:00:26.1 & +41:06:08.2\\
	position error ($3 \sigma$) & $1\,\farcs6$ & $13\,\farcs8$\\
	GC RA$^a$ (J2000) & 00:42:33.16 & 00:42:45.20\\
	GC Dec$^a$ (J2000) & +41:00:26.1 & +41:06:08.3\\
	distance to GC & $0\,\farcs5$ & $1\,\farcs1$\\
	kT [eV] & $48^{+2}_{-3}$ ($41^{+22}_{-18}$) & $74^{+32}_{-23}$\\ \noalign{\smallskip}
	\nh [\ohcm{21}] & $2.3\pm0.1$ ($2.3^{+3.0}_{-1.4}$) & $1.0^{+1.6}_{-0.9}$\\ \noalign{\smallskip}
	$L_{x}$ ($10^{38}$ \hbox{erg s$^{-1}$}) $^b$& $10.6 \pm 0.2$ ($20\pm4$) & 0.6 $\pm$ 0.1\\ \noalign{\smallskip}
	$L_{bol}$ ($10^{38}$ \hbox{erg s$^{-1}$}) & $28.7^{+0.2}_{-0.1}$ ($85\pm18$) & $1.0^{+0.7}_{-0.4}$\\ \noalign{\smallskip}
	$R$ (\power{9} cm) & $7.0^{+1.6}_{-0.7}$ ($15^{+28}_{-13}$) & $0.5^{+1.1}_{-0.3}$\\ \noalign{\smallskip}
\hline
\end{tabular}
\end{center}
\noindent
Notes:\hspace{0.3cm} $^a$: GC positions from \citet{2004A&A...416..917G}\\
\hspace*{1.1cm} $^b $: Unabsorbed; range 0.2 -- 1.0 keV.\\
\end{table}

Recently, \citet{2007ApJ...671L.121S} reported the very first nova found in a \m31 GC (\novagk). CNe in GCs are scarce, with just two sources known before 2007 that fit this definition. One was detected in the Galactic GC M\,80 whereas the second nova was found in a GC of the galaxy M\,87 \citep[see][and references therein]{2004ApJ...605L.117S}. According to \citet{2004ApJ...605L.117S} a third candidate (nova 1938 in the galactic GC M\,14) is less likely to be a CN. For all of this CNe there is no X-ray counterpart known.

\novag was found in Bol 111 on 2007 June 19.38, i.e. 141 days before our discovery of the SSS SS1 in Bol 111. Supersoft spectra like the one of SS1 are typical for X-ray counterparts of optical novae. Time lags of $\sim$150 days between an optical nova outburst and the first detection of the SSS counterpart have been observed before for other CNe \citep{2007A&A...465..375P}. Therefore, we identify the supersoft X-ray transient in the globular cluster Bol 111 with the CN \novagk.

No optical counterpart has been reported for SS2 in Bol 194. Therefore, we searched our optical monitoring data for indications of a nova outburst in the GC. These data are based on observations with the 45 cm ROTSE-IIIb telescope \citep{akerlof03} at the Turkish National Observatory (Bakirlitepe, Turkey), the robotic 60 cm telescope Livermore Optical Transient Imaging System \citep[Super-LOTIS,][]{2008AIPC.1000..535W} located at Steward Observatory (Kitt Peak, Arizona, USA), supplemented by archival data from K. Hornoch obtained at telescopes in  Lelekovice (35 cm telescope) and Ond\v{r}ejov (65 cm telescope).

Unfortunately, we did not discover a CN counterpart of SS2. However, due to seasonal gaps in the observation this does not rule out a CN in Bol 194. We conducted simulations of optical nova outbursts with different peak magnitudes and corresponding light curve decay times on any day between October 2004 and November 2007. These simulations took into account the intrinsic magnitude of the underlying GC. We find that most CN would have remained undetected in our optical data if they occurred during the periods 2005-02-15 -- 2005-05-12, 2006-02-19 -- 2006-05-09, or 2007-02-18 -- 2007-05-20, which is 990 -- 904, 621 -- 542, and 257 -- 166 days before the first detection of SS2 in X-rays. These detection gaps account for 23\% of the optical monitoring time. On the other hand, our coverage for 77\% of the time is almost complete, down to nova peak magnitudes of about 18.5 -- 19.0 mag in the R band.

If we assume that both SSSs are post-CNe and that the SSS phase lasts on average one year we find a nova rate of 0.015 yr$^{-1}$ GC$^{-1}$ for \m31. This is significantly larger than previous upper limits based on optical non-detections \citep[e.g. 0.005 novae yr$^{-1}$ GC$^{-1}$;][]{1992BAAS...24.1237T}. The connection of SSSs and CNe may therefore help to study the CN population in GCs.

%
%
\section{Very short SSS phase of \novaf}
\label{sec:novaf}
%
Our AO6 monitoring also led to the discovery of the very short SSS phase of the CN \novaf \citep{2009A&A...498L..13H}. We discovered \novaf in the optical as a candidate nova in our Super-LOTIS monitoring data of 2007 Nov 2.28 UT \citep{2007ATel.1257....1P} at RA = 00h42m37.29s, Dec = +41$\degr$17$\arcmin 10\,\farcs3$ (J2000, accuracy of 0.2"). The outburst date is well-constrained, since \novaf was not detected on the night before the discovery. In the second HRC-I observation of the \m31 X-ray monitoring campaign (starting 2007 Nov 17.76), we detected \novaf as a X-ray source. The X-ray position is in excellent agreement ($0\,\farcs1$) with the optical data. Thanks to our AO6 monitoring strategy, \novaf was visible in four consecutive \chandra observations before it faded and was not detected in the following \xmm observations. We computed HRC-I hardness ratios, as described in ``The Chandra Proposers Observatory Guide"\footnote{http://cxc.harvard.edu/proposer/POG/html/index.html;\hspace*{0.1cm} chapter 7.6}, from count rates in the bands S, M, and H (channels 1:100, 100:140, and 140:255). The ratios S/M = $-0.10\pm0.15$ and M/H = $0.09\pm0.15$ indicate a SSS spectrum with a $kT \lesssim 40\,\mbox{eV}$. This assumes an absorbed black body spectrum with an \nh $\geq$ 6.7\hcm{20}, the Galactic foreground absorption towards \m31 \citep{1992ApJS...79...77S}. Based on the indication of a SSS counterpart and the optical and X-ray light curves we classify \novaf as a CN.

\novaf was a very fast CN in the optical, exhibiting a very short SSS state in the X-ray with an appearance time of 6--16 days and a turn-off time of 45--58 days after the optical outburst. The appearance timescale implies an ejected mass of $(0.4-3)\times10^{-7}$\msun. To compute the ejected mass range we assume a typical value for the envelope expansion velocity of 2000 km s$^{-1}$, since we do not have an optical spectrum of \novaf, and that the SSS turns on when the absorbing hydrogen column density decreases to $\sim10^{21}$ cm$^{-2}$. This mass range is about two orders of magnitude lower than the ejected envelope masses of most \m31 novae discussed in \citet{2007A&A...465..375P}. The SSS turn off time constrains the burned mass to the range $(8-10)\times10^{-8}$\msun, according to \citet[][their Equation (5)]{2005A&A...439.1061S} Here we assume a bolometric luminosity of $3\times10^4L_{\odot}$ and a 
hydrogen mass fraction $X_H=0.5$. It is noteworthy that the burned mass is comparable to the ejected mass, within a factor of 2 -- 3. The extremely short SSS phase could have been caused either by a WD with $M_{WD} >$ 1.1 \msun\, for a standard hydrogen fraction in the envelope or by a very hydrogen-poor envelope in a WD with $M_{WD} \sim$ 1.0 \msun.

%
\section{Short time variations in the X-ray light curve of \novap}
\label{sec:novap}
%
A comprehensive analysis of our AO5 -- AO7 monitoring will be given in \citet{2009cHenze}. Here we present the short-time variability in the X-ray light curve of the CN \novapk.

\begin{figure}
	\resizebox{\hsize}{!}{\includegraphics[angle=270]{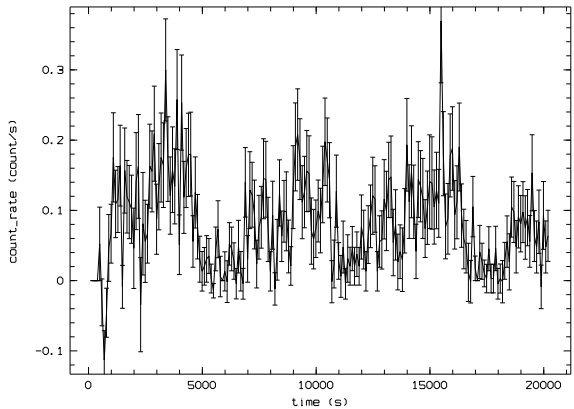}}
	\caption{Exposure and barycentre corrected \xmm EPIC PN light curve of \novap (0.2 -- 1.0 keV, 100s bins).}
	\label{fig:lc_novap}
\end{figure}

\novap was independently discovered in our optical \m31 monitoring \citep[with the Bradford Robotic Telescope Galaxy at the Tenerife Observatory; see][]{2006ATel..805....1P} and by K. Itagaki\footnote{see http://www.cfa.harvard.edu/iau/CBAT\_M31.html\#2006-04a}. The CN showed up as an X-ray source in only one \xmm observation on 2006-08-09 (ObsID 0405320601), which is 103 days after the first optical detection. The source is not detected on 2006-07-02 (\xmm ObsID 0405320501) and on 2006-09-30 (\chandra ObsID 7284), which are about 67 and 157 days after the optical discovery, respectively. \novap only shows photons with energies $<$ 0.7 keV and is therefore classified as a SSS. The EPIC PN light curve (see Fig.\,\ref{fig:lc_novap}) indicates significant variability with a period of about 2 hours.

In \m31, there were just three SSSs known previously which show periodicities in their X-ray light curves: the persistent SSS XMMU~J004252.5+411540 \citep[217.7s period;][]{2008ApJ...676.1218T}, the transient supersoft source 
XMMU~J004319.4+411758 \citep[865.5s period;][]{2001A&A...378..800O}, and the CN M31N~2007-12b (1100s period; see contribution of W. Pietsch in this volume). We want to mention that also in the Galaxy there are just a few novae known with periodicities in their light curves: RS~Oph \citep{2007ApJ...665.1334N}, V4743~Sgr \citep{2006MNRAS.371..424L}, and V1494~Aql \citep{2003ApJ...584..448D}.

The orbital periods of CVs in general are typically in the range of 1 -- 10 hours \citep{2003A&A...404..301R}. CNe have similar orbital periods \citep{2002AIPC..637....3W}. The pulsation and spin periods of WDs in CNe are typically shorter than 1 hour \citep[see e.g.][2500s pulsation period in nova V1494 Aql]{2003ApJ...584..448D}. The light curve variability of \novap might therefore be interpreted as the orbital period of the binary system.

%
\section{Summary}
\label{sec:discuss}
%
In this paper we review recent discoveries of SSSs in \m31. The discovery and good light curve coverage of fast transients like SS2 in Bol 194 or \novaf shows that our \xmm AO6/AO7 monitoring strategy was successful. Also for the AO5 source \novap a denser X-ray monitoring would have been useful for the study of light curve variability. In agreement with \citet{2007A&A...465..375P} we find that short supersoft states could play an important role in the SSS population of \m31.

\acknowledgements
We wish to thank the anonymous referee for helpful comments. The X-ray work is based in part on observations with \xmmk, an ESA Science Mission with instruments and contributions directly funded by ESA Member States and NASA. The \xmm project is supported by the Bundesministerium f\"{u}r Wirtschaft und Technologie / Deutsches Zentrum f\"{u}r Luft- und Raumfahrt (BMWI/DLR FKZ 50 OX 0001) and the Max-Planck Society. We would like to thank the \swift team for the scheduling of the ToO observations. M. Henze and H.S. acknowledge support from the BMWI/DLR, FKZ 50 OR 0405. G.S. acknowledges support from the BMWI/DLR, FKZ 50 OR 0405, and from grants AYA2008-04211-C02-01 and AYA2007-66256. M. Hernanz acknowledges support from grants AYA2008-01839 and 2009 SGR 315. A.R. acknowledges support through NASA/Chandra grant GO9-0024X. D. Hartmann acknowledges internal funding from Clemson University for partial support of the operation of Super-LOTIS.

\end{document}